\documentclass{appolb}
\usepackage{epsfig}
\usepackage{amsmath}
\usepackage{amssymb}
\begin{document}
\title{Economic and social factors in designing disease control strategies for epidemics on networks}

\author{A.~Kleczkowski\footnote{Electronic address: \texttt{adamk@mathbio.com}}, B. Dybiec\footnote{Electronic address: \texttt{bartek@th.if.uj.edu.pl}} and C.A.~Gilligan\footnote{Electronic address: \texttt{cag1@cam.ac.uk}} \address{$^{\dagger\S}$Dept.\ of Plant Sciences, University of Cambridge, Cambridge CB2 3EA, England\\$^\ddagger$Mark Kac Center for Complex Systems Research and Marian Smoluchowski Institute of Physics, Jagellonian University, 30--059 Krak\'ow, Poland}}
\maketitle

\begin{abstract}

Models for control of epidemics on local, global and small-world
networks are considered, with only partial information accessible
about the status of individuals and their connections. The main goal
of an effective control measure is to stop the epidemic at a lowest
possible cost, including treatment and cost necessary to track the
disease spread. We show that delay in detection of infectious
individuals and presence of long-range links are the most important
factors determining the cost. However, the details of long-range
links are usually the least-known element of the social interactions
due to their occasional character and potentially short life-span.
We show that under some conditions on the probability of disease
spread, it is advisable to attempt to track those links. Thus,
collecting some additional knowledge about the network structure
might be beneficial to ensure a successful and cost-effective
control.

\end{abstract}


\maketitle

\section{Introduction}\label{sec-intro}

One of the main goals of epidemiological modeling is to provide
guidelines for controlling disease outbreaks. Traditionally this has
been understood in terms of reducing the number of infected
individuals. With a cheap vaccination available, `blind' vaccination
of a large proportion of individuals might be a simple and yet
optimal solution \cite{dybiec1}. However, in many cases the epidemic
must be stopped at a manageable cost and with potentially limited
resources, leading to a mixture of preventive and responsive
measures. In the simplest case the goal of a successful prevention
and eradication programme is to minimize a number of individuals who
have either been treated or have been through the infection.

In a series of previous papers \cite{dybiec1,dybiec} we have studied
the suitability of local control strategies for stopping spread of
diseases on networks with a mixture of local and global links. These
include `small-world' networks \cite{small}, with a majority of
contacts between nearest neighbors and a small number of global
links. By a local strategy we understand control measures limited to
some neighborhood of an infected individual. We have proposed a
strategy that is a mixture of responsive and preventive actions. A
control event is triggered by an appearance of a symptomatic
individual (responsive measure) and spans not only this individual
but also its immediate neighbors on a certain control network
(preventive measure).

The preventive control (analogous to a ring-vaccination strategy) is
necessary because of the delay between the onset of infectiousness
of an individual and the onset of symptoms. Thus, there is a
possibility of pre-symptomatic yet infectious individuals to be
located in the neighborhood of the observed disease case. The
preventive local control strategy attempts to treat such potential
cases. The crucial assumption in our paper is that the network that
defines the control neighborhood is only a subset of the network on
which the disease spreads and in particular contains only a number
of long-range links. This reflects the limited ability of medical
authorities to track and follow contacts between individuals leading
to spread of the disease. In particular, we ask the following
question: how detailed should our knowledge be of the network
structure to be able to stop the disease at the lowest possible
cost? We compare different strategies by looking at the final size
of the epidemics including individuals who have been through the
disease as well as those treated \cite{dybiec}. We also include an
additional cost associated with tracking long-range links.

\section{Model}\label{sec-model}

The model of epidemic spread and the associated control must take
into account the topology of the network on which the epidemic
spreads, the topology of the sub-network which is used for control,
the state of each individual and transitions between different
states. We consider two basic topological structures, a
1-dimensional small world topology (SW1D) \cite{small} and
2-dimensional small world topology (SW2D). The disease spreads on
the full network, including local and global links. The control
measures can only follow a subset of those links and in particular
for the SW1D and SW2D topologies we assume that the subset contains
all local links and a number of additional shortcuts. This approach
is caused by the fact that it is much easier to track local
interactions, interactions with surrounding individuals, fields and
farms, than long-range interactions, which might be caused by
geographical, technical, cultural or economical factors.

The epidemiological part of the model is based on an SIR model
\cite{anderson} modified so that it includes pre- and
post-symptomatic individuals (who can both contribute to the spread
of the infection) and recovered as well as treated individuals.

\subsection{Topology}\label{subsec-topology}
\begin{figure}[!h]
\begin{center}
\includegraphics[angle=0, width=8.0cm, height=5.0cm]{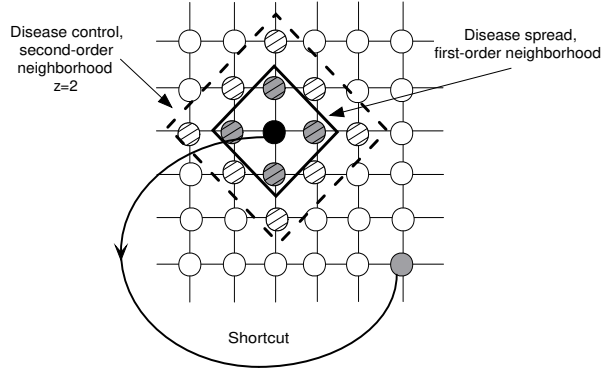}
\caption{SW2D topology: In this example, a detected individual
(black circle) is in contact with its four nearest-neighbors on the
disease network and to one node connected by a shortcut (gray
circles to indicate non-symptomatic infected individuals). The
control might then be applied locally, limited to the eight
second-order neighbors and individual itself on a treatment network
(marked by a square). In general a given ratio of additional
shortcuts can be incorporated in the disease control neighborhood
making control more efficient.} \label{model-topology}
\end{center}
\end{figure}

SW2D topology is constructed from a regular lattice, with periodic
boundary conditions, to which a given number of additional random
shortcuts is added. Thus, every individual placed on the SW2D
topology interacts with its four nearest neighbors and some other
individuals via additional shortcuts (Fig.\ \ref{model-topology}).
The SW1D topology is constructed in a similar way, by adding
long-range links to a one-dimensional ring. For compatibility
between SW2D and SW1D topology every node of the initial ring has 4
first order neighbors, 2 of them located on the left hand side and 2
on the right hand side of the given node.

In the first instance, the control network contains the regular
(local) part of the infection network. In addition, we assume that a
certain number of long-range links is included in the control
network. This reflects an additional effort that a government or
health quthority put into disease tracking. A control neighborhood
of given order, $z$, is constructed in an iterative way. Starting
from the infected node the first order control neighbors are
localized. The second order neighbors are then found as first order
neighbors of the previous group. The whole procedure is repeated
iteratively $z$ times. A single control action involves all
individuals in the control neighborhood of order $z$.

\subsection{Individuals and Transitions}\label{subsec-trans}
Individuals are placed on a given topology and can be in one of the
following states:
\begin{enumerate}
\setlength{\topsep}{0pt} \setlength{\parsep}{0pt}
\setlength{\itemsep}{2pt} \setlength{\parskip}{0pt}
\item{$\mathbf{S}$ -- susceptible (or healthy), which can be infected with probability $p$ by any infectious or detected individual in its epidemic neighborhood;}
\item{$\mathbf{I}$ -- infectious (infected but pre-symptomatic); can infect other nodes from its epidemic neighborhood but cannot trigger a control measure. In addition, with probability $q$ it can spontaneously move to the detected class, i.e. symptoms become observable;}
\item{$\mathbf{D}$ -- detected (infected and symptomatic), can infect other nodes from its epidemic spread neighborhood. In addition, it can spontaneously move to the recovered class (with the probability $r$) or can trigger a treatment
measure with the probability $v$ that includes all individuals
within its control neighborhood;}
\item{$\mathbf{R}$ -- recovered. This class includes individuals that have been through the disease, can be treated but cannot become re-infected;}
\item{$\mathbf{V}$ -- vaccinated (treated). Individuals in this class have been in a control neighborhood of a detected individual when the treatment event was triggered. They cannot become re-infected.}
\end{enumerate}

We assume that all nodes in the network are occupied. The initial
state is a mixture of a majority of susceptible individuals with an
addition of few ($0.1\%,0.5\%$ or $5\%$) infectious (symptomatic)
individuals. We denote the total number of nodes by $N$ and the
number of susceptible nodes by $S$, infected by $I$, detected by
$D$, recovered by $R$ and treated (vaccinated) by $V$.

\subsection{Simulations}\label{subsec-sim}

Details of the simulations are given in \cite{dybiec,dybiec1}. The
model was updated synchronously and the simulation loop was
performed until the number of infected individuals was equal to
zero, i.e. until $\mathrm{T_{max}}$ such that
$I(\mathrm{T_{max}})+D(\mathrm{T_{max}})=0$. In every iteration,
spontaneous transitions from $\mathbf{I}\to\mathbf{D}$,
$\mathbf{D}\to\mathbf{R}$ and state-dependent transitions
$\mathbf{S}\to\mathbf{I}$, $\mathbf{D}\to\mathbf{V}$ were performed.

We consider three treatment strategies, random vaccination, local
vaccination and a mixed strategy combining local vaccination with
tracking of long-range links. In the random `blind' vaccination, the
given ratio of randomly chosen individuals is vaccinated shortly
after the first detection of the disease. For local treatment all
individuals up to a given order $z$ surrounding and including the
detected infected individual, are vaccinated regardless of their
current disease status. For the mixed strategy, a certain proportion
of long-range links is also tracked and individuals to which the
detected individual is linked are treated as well.

For a given set of parameters the simulation was averaged over 50
realizations for the total number of nodes equal to 2500 (i.e. the
SW2D topology is created from the square $50\times50$ lattice), with
or without addition of a fixed number of 1023 long-range links.
Larger sizes of the networks and larger number of realizations were
explored as well, but they did not improve or change the results.

\section{Results}\label{sec-results}

Simulation results were analyzed to extract information that is
relevant for the design of an optimal control strategy. In
particular, we look at a severity index, a combined number of
treated and recovered individuals, $X\equiv R(\infty)+V(\infty)$ at
the end of an epidemic. This quantity represents the combined
severity of an untreated epidemic, $R(\infty)$, and the costs of
treating it, $V(\infty)$. In this paper, we mainly focus on effects
of a network structure (including shortcuts) and probability of
spread, $p$, on the severity index, $X$, of the epidemic and the
optimal extent of a control neighborhood $z_c$. $z_c$ is defined as
such a diameter of control neighborhood for which $X(z_c)$ is
minimal. All other parameters, except $r$ which was set to $0.01$,
take all possible values from the allowed domains and
$z\in\{1,2,\dots,15\}$. In addition, we vary the structure of the
control network by changing the proportion
($T_L=\{0\%,10\%,20\%,\dots,100\%\}$) of long-range links
(shortcuts) that are tracked and included in the control
neighborhood.

\begin{figure}[!h]
\begin{center}
\includegraphics[angle=0, width=12.0cm, height=8.0cm]{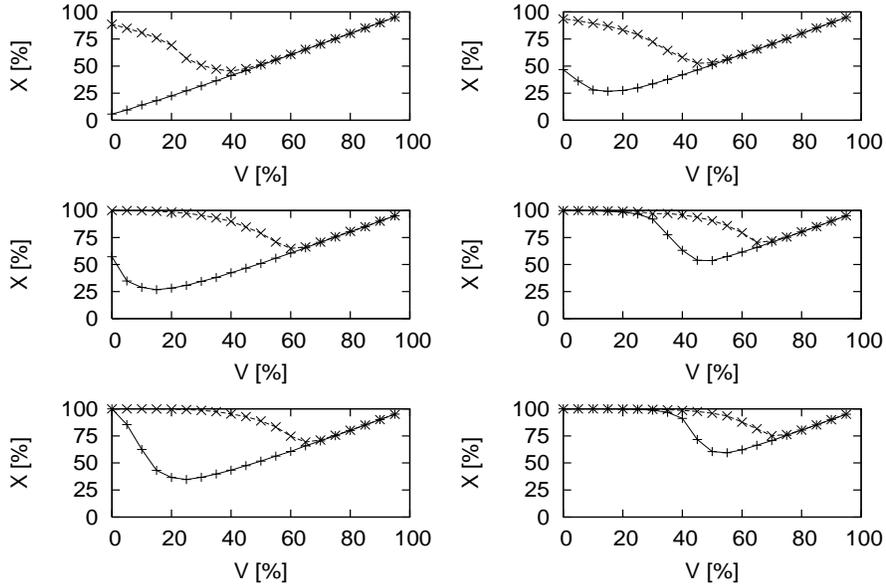}
\caption{$X=R(\infty)+V(\infty)$ as a function of proportion of
initially vaccinated individuals for SW1D topology (left panel) and
SW2D topology (right panel) for various values of infection
probability, $p$: $p=0.01$ (top panel), $p=0.05$ (middle panel) and
$p=0.5$ (lower panel). Other parameters: $q=0.5$, $r=0.01$.
Different symbols correspond to various numbers of additional
shortcuts: `$+$' 0 shortcuts, `$\times$' 1023 shortcuts. Initially,
at $t=0$, 0.5\% of all individuals were in the symptomatic class.}
\label{blindvaccinationcost}
\end{center}
\end{figure}

We first consider a `blind' vaccination strategy,
Fig.~\ref{blindvaccinationcost} and the effect of different
proportions of vaccinated individuals is assessed on the impact of
disease. This strategy is effective when applied early and the
number of non-local links is small, see
Fig.~\ref{blindvaccinationcost}. Addition of long-range links or
delaying the application of the `blind' treatment renders it
ineffective, \cf Fig.~\ref{blindvaccinationcost}. In addition, from
the social point of view, such a strategy is difficult to accept,
because it is purely preventive and control measures are focused
only on initial vaccination of randomly chosen individuals without
any further actions during the outbreak.

\begin{figure}[!h]
\begin{center}
\includegraphics[angle=0, width=8.0cm]{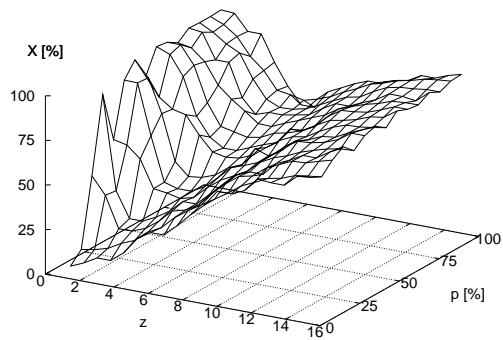}
\caption{$X\equiv R(\infty)+V(\infty)$ as a function of the
infection probability $p$ and diameter of the vaccination $z$ for
SW2D network with 63 additional shortcuts. Other parameters:
$q=0.5$, $v=0.1$ and $r=0.01$. Initially, at $t=0$, 0.5\% of all
individuals were in the symptomatic class.} \label{optimal_rl63_3d}
\end{center}
\end{figure}

The next group of possible control strategies is characterized by a
mixture of responsive and preventive actions. As new foci of the
disease are formed and spread, they trigger control measures that
are applied in a broader neighborhood of detected symptomatic
individuals. The extended control neighborhood compensates for the
lack of our knowledge about the exact state of individuals and the
exact structure of interactions. The severity index $X\equiv
V(\infty)+R(\infty)$ is plotted in Fig.~\ref{optimal_rl63_3d} as a
function of infection probability $p$ and the control neighborhood
size, $z$. For each value of $p$, there exist an optimal value $z_c$
for which the control measures are most efficient. If $z<z_c$ the
disease escapes the control, while for $z>z_c$ too many individuals
are vaccinated. The exact shape of the surface depends on network
properties and epidemic parameters. Nonlocal interactions make
minima less pronounced; nevertheless, purely local strategies are
capable of stoping epidemics even in the presence of long-range
links \cite{dybiec}.

\begin{figure}[!h]
\begin{center}
\includegraphics[angle=0, width=12.0cm, height=8.0cm]{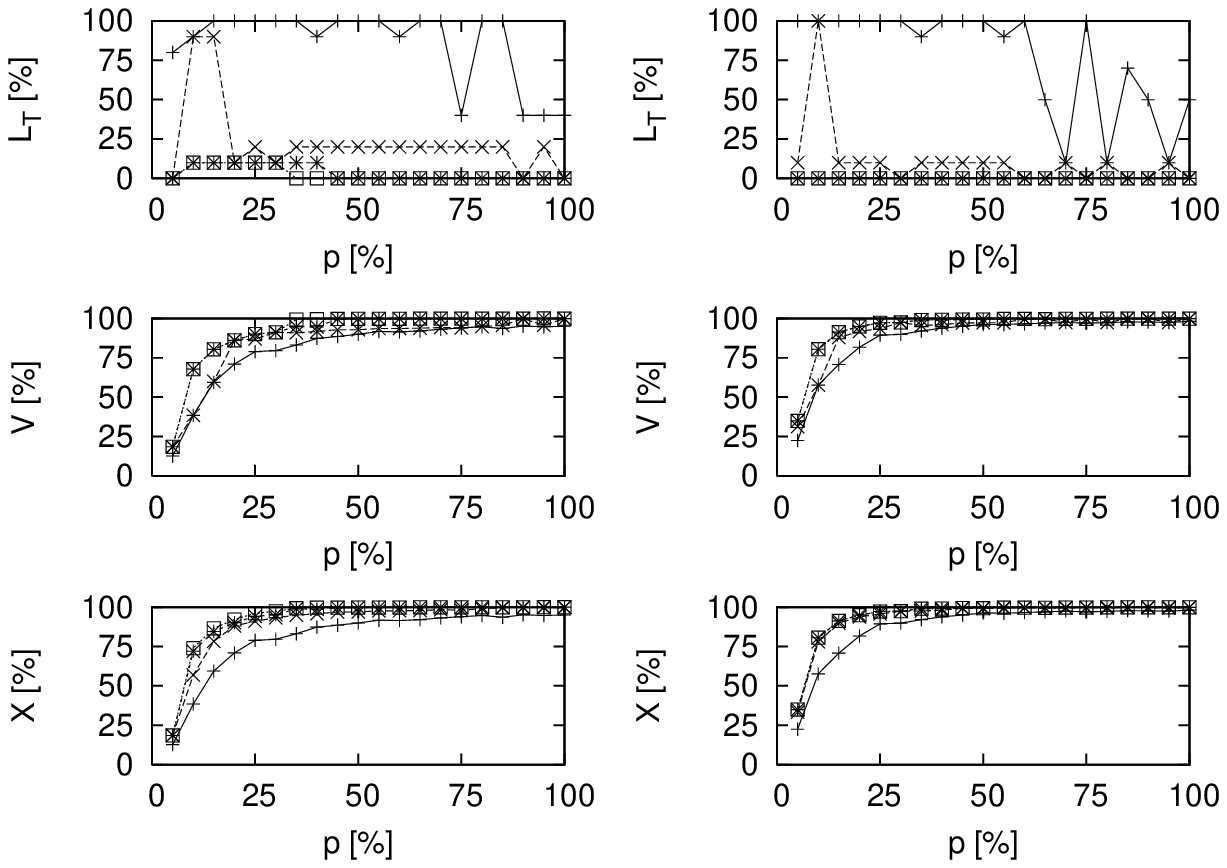}
\caption{Ratio of the tracked links $L_T$ (top panel), proportion of
the vaccinated individuals $V(\infty)$ (middle panel) and $X\equiv
R(\infty)+V(\infty)+\alpha\cdot L_T$ (bottom panel) as a function of
the infection probability $p$ for SW1D topology (left panel) and
SW2D topology (right panel). Other parameters: $q=0.1$, $v=0.1$ and
$r=0.01$. Initially, at $t=0$, 0.5\% of all individuals were in the
symptomatic class. Different symbols correspond to various cost of a
single non-local link tracking $\alpha'$: `$+$' 0 , `$\times$' 0.5,
`$\ast$' 1.0, `$\square$' 1.5.} \label{2d_v01_q01_005}
\end{center}
\end{figure}

\begin{figure}[!h]
\begin{center}
\includegraphics[angle=0, width=12.0cm, height=8.0cm]{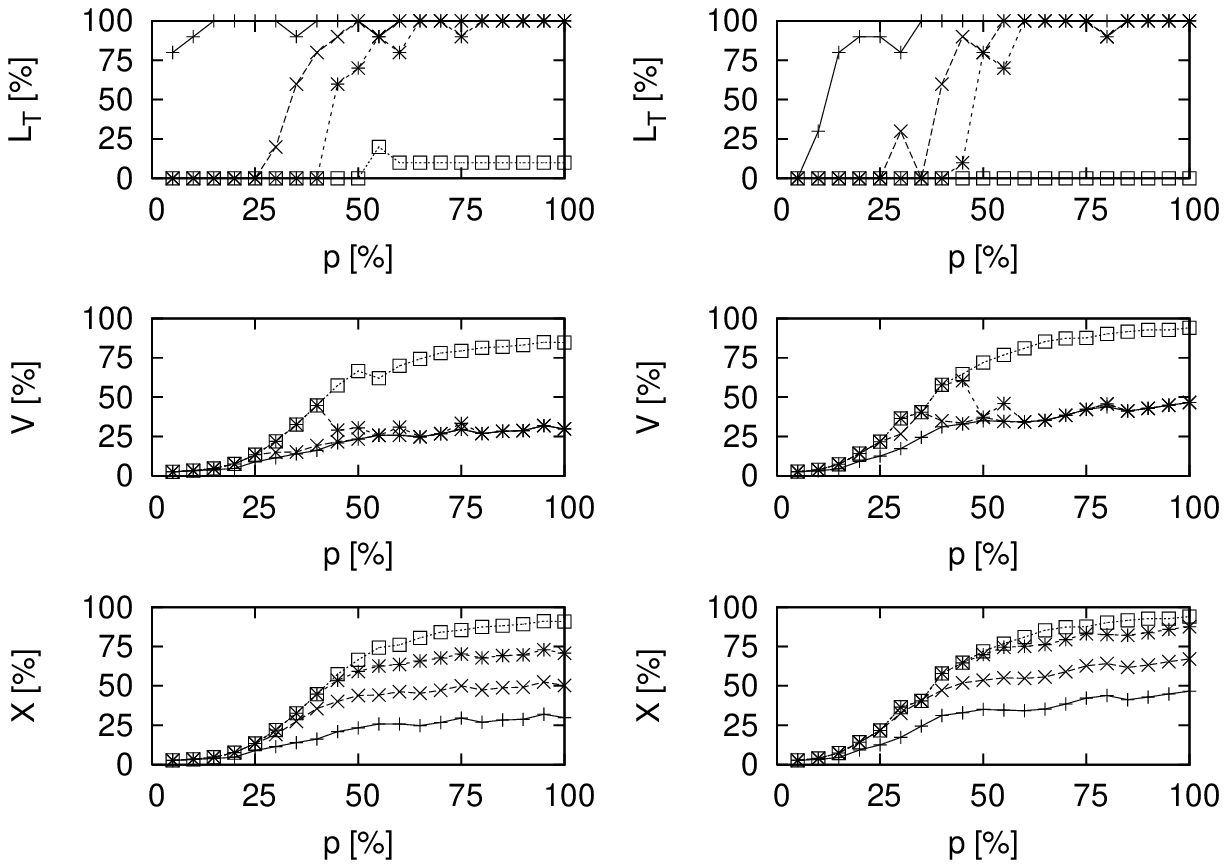}
\caption{The same as in Fig.~\ref{2d_v01_q01_005} for $q=0.5$,
$v=0.5$.} \label{2d_v05_q05_005}
\end{center}
\end{figure}


Epidemics can spread not only to the nearest neighbors but also, via
random non-local shortcuts, to distant part of the network. On the
one hand, long-range links are crucial for the spread of the
outbreak. On the other hand they are hard to identify and their
identification requires an additional cost. Therefore, for knowledge
oriented strategies, more general cost functions need to be
considered. We propose $X\equiv V(\infty)+R(\infty)+\alpha\cdot
L_T$, where $L_T$ represents the ratio of identified to the total
number of shortcuts and $\alpha$ is the cost associated with contact
tracking.

Figs.~\ref{2d_v01_q01_005}--\ref{2d_v05_q05_005} show ratio of
tracked links (top panel), the number of vaccinated individuals
(middle panel) and cost function $X$ (lower panel), corresponding to
the optimal solutions. In the following we examine the influence of
incubation time, controlled by $q$, and effectiveness of the
vaccination, $v$, on the optimal strategy.

For the parameters used in this paper, the cost associated with an
optimal strategy is generated mainly by vaccination and links
tracking. The relative importance of these two factors depends on
the cost of tracking a single long-range link,
$\alpha'=\alpha/1023$. When links tracking is cheap, it is optimal
to track all shortcuts, see
Figs.~\ref{2d_v01_q01_005}--\ref{2d_v05_q05_005} (top panel). When
disease incubation time is long (small $q$) and vaccination is
inefficient (small $v$) detailed contact tracking is less important
and costs are largely influenced by treatment, \cf
Fig.~\ref{2d_v01_q01_005}. The combined effect of the long
incubation time and low effectiveness of vaccination decrease the
effect the additional knowledge about long range links has on the
control. When the incubation is time long, epidemics can infect
large proportion of individuals before they are detected. For short
incubation times (large $q$) and more effective treatment (large
$v$), there is a clear distinction between strategies applying
contact recognition and purely local strategies, \cf
Fig.~\ref{2d_v05_q05_005}. The recognition of shortcuts, despite the
associated costs, can significantly decrease the number of
individuals that need to be vaccinated to eradicate epidemics.
Furthermore, such strategies lead to smaller value of the severity
index $X$ than purely local strategies, \cf middle and bottom panels
of Fig.~\ref{2d_v05_q05_005}.

\section{Discussion}\label{sec-discussion}

Designing control strategies for networks incorporating long-range
links is complicated. In the simplest case, we envisage treating
infected and/or susceptible individuals so that the disease progress
is slowed down or even stopped. Examples of such treatment include
preventive vaccination, culling of animals and quarantine of fields
or individuals. For networks with only short-range interactions the
spread of a disease is geographically limited and can therefore be
contained locally \cite{dybiec}. For non-local networks there is
always a possibility of infection jumping to another location to
form a new focus. In designing control strategies for such networks
it is necessary to know not only the geographical location of new
cases (so that they and their immediate neighbors can be treated)
but also all possible connections that can span the whole
population. Obtaining this information can be very expensive and
time consuming. With the authorities faced by a large-scale epidemic
the collection of such data might be difficult and might lead to
many inappropriate decisions. It is thus imperative to use
epidemiological models to explore the possibilities of simplifying
the control strategies.

Most models of disease spread used to predict its advance and to
design efficient control measures assume a perfect knowledge of both
the status of each individual (healthy vs. infectious) and the
network structure (who acquires the disease from whom
\cite{anderson,longrangevacc}). Among the epidemiological
parameters, the difference between the onset of infectiousness and
the earliest detectability of the disease is the key issue for
controlling the disease. For most diseases an individual can be
infectious even though the infection cannot be detected and
controlled. Such an individual can be a source of further infections
for a relatively long time until the source is identified and
controlled by isolation or treatment. In many cases, even
post-symptomatic individuals cannot be treated straight after the
detection, which further adds to a spread of the epidemic. Control
strategies should aim at decreasing the time until control measures
are applied by increasing detectability and speeding up control.

We have shown that long-range links dramatically reduce the
effectiveness of local control measures. Our results show that in
some cases it is possible to control epidemics with only limited
knowledge about interactions between individuals. If this is not
possible, our model gives guidance on conditions under which it is
advisable to attempt to track long-range links, despite the high
costs associated with such a strategy. From the economic point of
view, contact tracking is important when disease incubation time is
short and vaccination is efficient. Furthermore, if the epidemics is
highly infectious, knowledge oriented strategies lead to a
significant decrease in the severity index characterizing the costs
of disease eradication.

There is a clear distinction between the case when the control
measure works and when it does not. If the control neighborhood is
too small, or we track insufficient numbers of long-range links, the
disease keeps escaping the treatment and as a result we need to
treat practically the whole population. Making the ring of control
even a fraction larger might lead to a dramatic increase in the
efficiency of the control strategy. Similarly, incorporating more
long-range links might improve the effectiveness of the control
measures.

The research was initiated under the British Council -- Polish State
Committee for Scientific Research (KBN) grant WAR 342/01. B.D. was
supported by the Polish State Committee for Scientific Research
(KBN) grant 2P03B0 8225 (2003--2006) and by the Foundation for
Polish Science through the domestic grant for young scientists
(2005). A.K. was supported by DEFRA and C.A.G. by BBSRC.

\end{document}